\documentclass[10pt,a4paper,twocolumn,superscriptaddress]{revtex4}

\usepackage{textcomp}
\usepackage{color}
\usepackage{amsmath}
\usepackage{mathrsfs}
\usepackage{graphicx}
\usepackage{subfig}
\usepackage{epsfig}
\usepackage{hyperref}

\usepackage[squaren,Gray]{SIunits}
\usepackage{pdfpages}
\usepackage{float}
 \usepackage{amssymb}

\begin{document}
\definecolor{c_red1}{RGB}{148,0,0}
\definecolor{c_blue1}{RGB}{6,77,135}

\captionsetup[figure]{justification=raggedright}


\title{Parametrically driven Kerr cavity solitons}

\author{Nicolas Englebert}
\email{nicolas.englebert@ulb.ac.be}
\affiliation{Service OPERA-\textit{photonique}, Universit\'e libre de Bruxelles (U.L.B.), 50~Avenue F. D. Roosevelt, CP 194/5, B-1050 Brussels, Belgium}

\author{Francesco De Lucia}
\affiliation{Service OPERA-\textit{photonique}, Universit\'e libre de Bruxelles (U.L.B.), 50~Avenue F. D. Roosevelt, CP 194/5, B-1050 Brussels, Belgium}
\affiliation{Optoelectronics Research Centre, University of Southampton, SO17 1BJ, United Kingdom}

\author{Pedro Parra-Rivas}
\affiliation{Service OPERA-\textit{photonique}, Universit\'e libre de Bruxelles (U.L.B.), 50~Avenue F. D. Roosevelt, CP 194/5, B-1050 Brussels, Belgium}

\author{Carlos Mas Arabí}
\affiliation{Service OPERA-\textit{photonique}, Universit\'e libre de Bruxelles (U.L.B.), 50~Avenue F. D. Roosevelt, CP 194/5, B-1050 Brussels, Belgium}

\author{Pier-John Sazio}
\affiliation{Optoelectronics Research Centre, University of Southampton, SO17 1BJ, United Kingdom}

\author{Simon-Pierre Gorza}
\affiliation{Service OPERA-\textit{photonique}, Universit\'e libre de Bruxelles (U.L.B.), 50~Avenue F. D. Roosevelt, CP 194/5, B-1050 Brussels, Belgium}

\author{Fran\c{c}ois Leo}
\affiliation{Service OPERA-\textit{photonique}, Universit\'e libre de Bruxelles (U.L.B.), 50~Avenue F. D. Roosevelt, CP 194/5, B-1050 Brussels, Belgium}

\begin{abstract}
Temporal cavity solitons are optical pulses that propagate indefinitely in nonlinear resonators~\cite{wabnitz_suppression_1993,leo_temporal_2010,herr_temporal_2014}. They are currently attracting a lot of attention, both for their many potential applications and for their connection to other fields of science. 
Cavity solitons are phase locked to a driving laser. This is what distinguishes them from laser dissipative solitons~\cite{grelu_dissipative_2012} and the main reason why they 
are excellent candidates for precision applications such as optical atomic clocks~\cite{newman_architecture_2019}.
To date, the focus has been on driving Kerr solitons close to their carrier frequency, in which case a single stable localised solution exists for fixed parameters~\cite{wabnitz_suppression_1993}.
Here we experimentally demonstrate, for the first time, Kerr cavity solitons excitation around twice their carrier frequency. In that configuration, called parametric driving, two solitons of opposite phase may coexist~\cite{longhi_ultrashort-pulse_1995}. We use a fibre resonator that incorporates a quadratically nonlinear section and excite stable solitons by scanning the driving frequency. 
Our experimental results are in excellent agreement with a seminal amplitude equation~\cite{longhi_hydrodynamic_1996}, highlighting connections to hydrodynamic~\cite{wu_observation_1984,miles_parametrically_1984} and mechanical systems~\cite{denardo_observations_1992}, amongst others~\cite{bondila_topography_1995}.
Furthermore, we experimentally confirm that two different phase-locked solitons may be simultaneously excited and harness this multiplicity to generate a string of random bits, thereby extending the pool of applications of Kerr resonators to random number generators~\cite{marandi_all-optical_2012} and Ising machines~\cite{inagaki_coherent_2016}. 
\end{abstract}

\maketitle

\noindent 
\begin{figure*}
\includegraphics[scale=0.9]{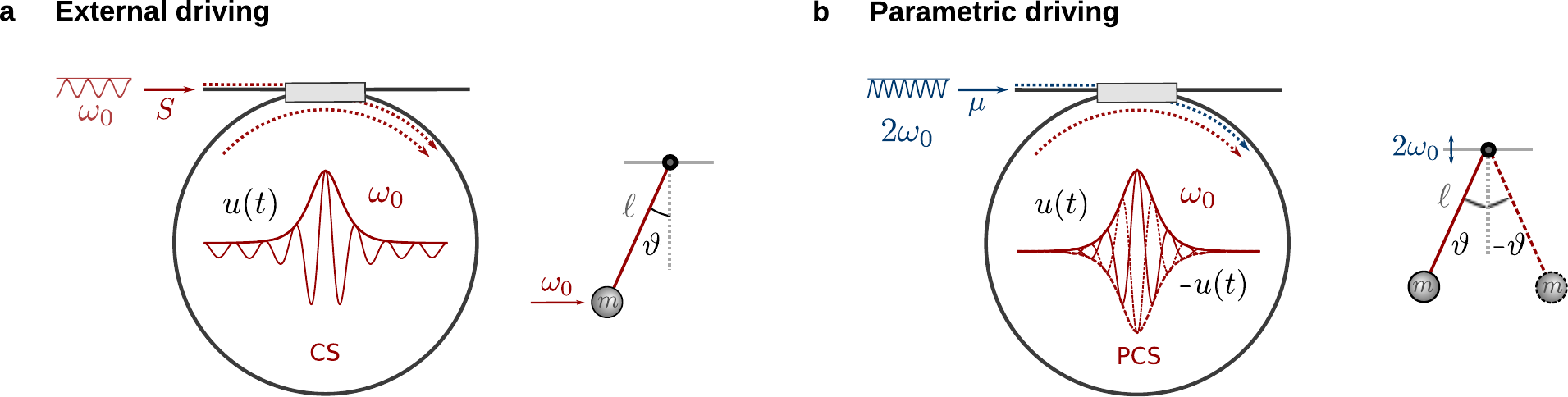}
\label{fig:concept}

\begin{center}
\begin{tabular}{p{8.5cm}p{0cm}p{9cm}}
  \begin{equation}\label{eq:ENLSE}
    \hspace{-1cm}\dfrac{\partial u}{\partial T}=
\left(-1+i(|u|^2-\Delta)+i\dfrac{\partial^2}{\partial \tau^2}\right)u + {\color{c_red1}S}
  \end{equation}
  &&
  \begin{equation}\label{eq:PNLSE}
        \dfrac{\partial u}{\partial T}=
\left(-1+i(|u|^2-\Delta)+i\dfrac{\partial^2}{\partial \tau^2}\right)u + {\color{c_blue1}\mu} u^*
  \end{equation} 
\end{tabular}
\end{center}
\vspace{-0.5cm}
\caption{\textbf{Illustration of the differences between external and parametric driving.} {\bf a},\,Schematic of an optical Kerr resonator pumped close to its natural response frequency. Solitons are solutions of externally driven nonlinear Schrödinger equation \eqref{eq:ENLSE}. Akin to a externally driven damped pendulum, a single phase-locked solution exists. {\bf b},\,Schematic of an optical Kerr resonator pumped close to twice its natural response frequency. Solitons are solutions of the parametrically driven nonlinear Schrödinger equation \eqref{eq:PNLSE}. Akin to a parametrically driven damped pendulum, two different phase-locked solutions exist.} 
\label{fig:concept}
\end{figure*} 
The spontaneous formation of patterns is encountered across many fields of science.
Spatially extended nonlinear systems may be brought away from equilibrium, where spatiotemporal patterns emerge~\cite{cross_pattern_1993}.
Examples include convection rolls in heated fluids~\cite{ahlers_heat_2009}, vegetation patches in arid regions~\cite{lejeune_localized_2002}, as well as localised structures in vibrated layers of sand~\cite{umbanhowar_localized_1996}.
These complex patterns can often be described by relatively simple reaction/diffusion equations that capture most of the nonlinear dynamics~\cite{cross_pattern_1993}. These so called amplitude equations have been shown to be universal. Very different systems in terms of microscopical physical laws can, under some conditions, be governed by the same macroscopic equation, providing important connections between distinct fields of science.

One such class of equations are the dissipative nonlinear Schrödinger equations (NLSE) which describe pattern formation in charge density condensates, driven plasmas, surface waves and optical resonators amongst others (see \cite{barashenkov_existence_1996,bondila_topography_1995} and references therein).
The conservative NLSE admits exact solitary wave solutions~\cite{zakharov_exact_1972} and similar localised structures can be found when dissipation and forcing are added to the system. 
As in one-dimensional oscillators such as the driven pendulum, the forcing can be external~\cite{barashenkov_existence_1996} or parametric~\cite{bondila_topography_1995}. In the former, the energy is transferred by exciting the systems close to its natural frequency. In the latter, the energy is injected by periodically varying a parameter of the system at twice the system's response frequency.
Parametric forcing of spatially extended systems has been intensely studied since the first reports, by Michael Faraday, of patterns on a vibrating surface~\cite{faraday_peculiar_1831}. Parametrically driven NLSE solitons have been reported in hydrodynamics~\cite{wu_observation_1984} and in chains of oscillators~\cite{denardo_observations_1992}, and have been predicted to exist in optical resonators in 1995~\cite{longhi_ultrashort-pulse_1995}. They constitute a subclass of optical dissipative solitons~\cite{grelu_dissipative_2012} along with temporal cavity solitons of the externally forced NLSE~\cite{wabnitz_suppression_1993,leo_temporal_2010,herr_temporal_2014}, which have been shown to underpin the formation of ultra-coherent optical frequency combs \cite{coen_modeling_2013,parra-rivas_dynamics_2014}. Note that temporal cavity solitons are commonly referred to as dissipative Kerr solitons in the context of microresonators~\cite{kippenberg_dissipative_2018}.

The main differences between externally forced cavity solitons (CSs) and parametrically driven cavity solitons (PCSs) are illustrated in Fig.~\ref{fig:concept}. 
CSs are solutions of the well known externally driven NLSE~\eqref{eq:ENLSE}, often called the Lugiato-Lefever equation~\cite{lugiato_spatial_1987,haelterman_dissipative_1992}. They sit on a homogeneous background and a single phase locked solution exists for fixed detuning and driving power~\cite{wabnitz_suppression_1993,nozaki_chaotic_1985}.
PCSs, on the other hand, are solutions of the parametrically driven NLSE~\eqref{eq:PNLSE}.
They lack a homogeneous background and two stable solutions, of opposite phase, may coexist~\cite{miles_parametrically_1984,longhi_ultrashort-pulse_1995}. 
This multiplicity opens up several new avenues for soliton coalescence~\cite{wang_universal_2017,cole_soliton_2017} as already demonstrated in hydrodynamics~\cite{wu_observation_1984}.
Moreover, stable optical pulses of opposite phases can be used to implement random number generators~\cite{marandi_all-optical_2012} and Ising machines~\cite{inagaki_coherent_2016}. 
Here, we report the first experimental characterisation of the parametric Kerr cavity soliton. We implement an all-fibre singly resonant degenerate optical parametric oscillator (OPO), which is well described by the parametric nonlinear Schrödinger equation (PNLSE). We measure backgroundless sech-shaped optical waves and show that solutions with different phases may coexist in the resonator. As a proof of principle experiment, we generate a short series of random numbers using PCSs.\newpage

\begin{figure}[H]
\includegraphics[scale=.68]{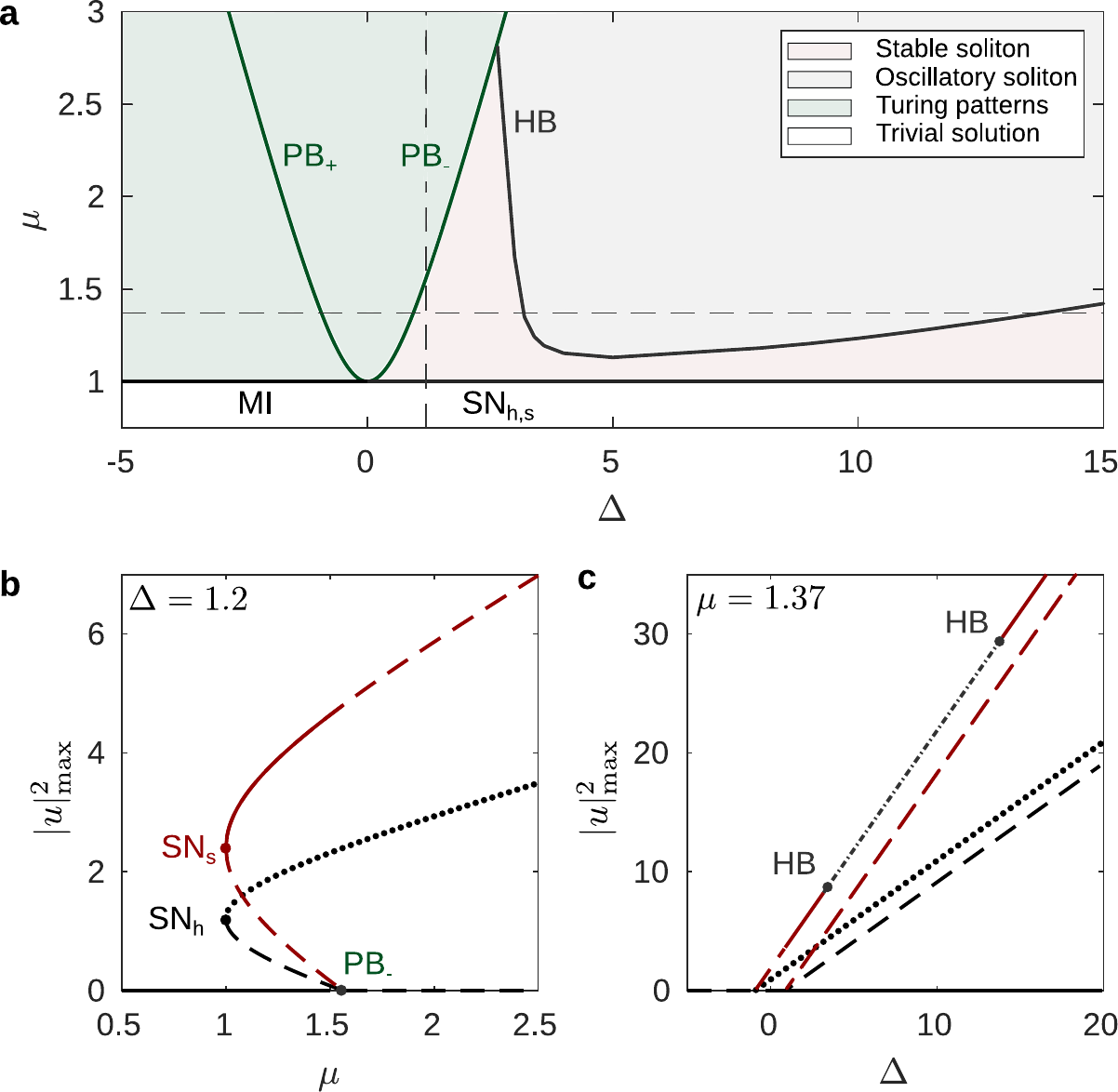}
\caption{\textbf{Bifurcation structure of the PNLSE.} {\bf a},\,Phase diagram in the $(\Delta,\mu)$-parameter space showing the main dynamical regions of the system. The bifurcation lines are the pitchfork bifurcations (PB$_\pm$) corresponding to the degenerate OPO threshold (green solid line) and the Hopf bifurcation (HB) line. The black line at $\mu=1$ corresponds to the saddle-node bifurcation of both the non-trivial homogeneous state SN$_{h}$ and the soliton state SN$_s$ for $\Delta>0$, and to modulation instability for $\Delta<0$. {\bf b},\,Bifurcation diagram showing the soliton branches (red line) as well as the homogeneous states (black line) as a function of $\mu$ for $\Delta=1.2$. The solid lines corresponds to stable states, dashed lines to homogeneously unstable states  and the dotted line to modulationnaly unstable states. {\bf c},\,Bifurcation diagram as a function of $\Delta$ for $\mu=1.37$. The soliton develops breathing behaviour in-between the HBs (see dotted-dashed line).}
\label{fig:Bifurcation}
\end{figure} 
OPOs are staples of nonlinear optics but the bulk of their usage has been in the homogeneous regime for frequency translation~\cite{myers_quasi-phase-matched_1995}. Recently, in the context of frequency comb formation, there has been interest in pattern formation in continuous wave pumped OPOs through cascaded three wave mixing ~\cite{mosca_modulation_2018,bruch_pockels_2020}. 
Our work extends the applications of OPOs by showing that they may host phase-locked Kerr solitons.

\noindent\textbf{Bifurcation analysis}\\
\begin{figure*}
  \centerline{\includegraphics[scale=1.0]{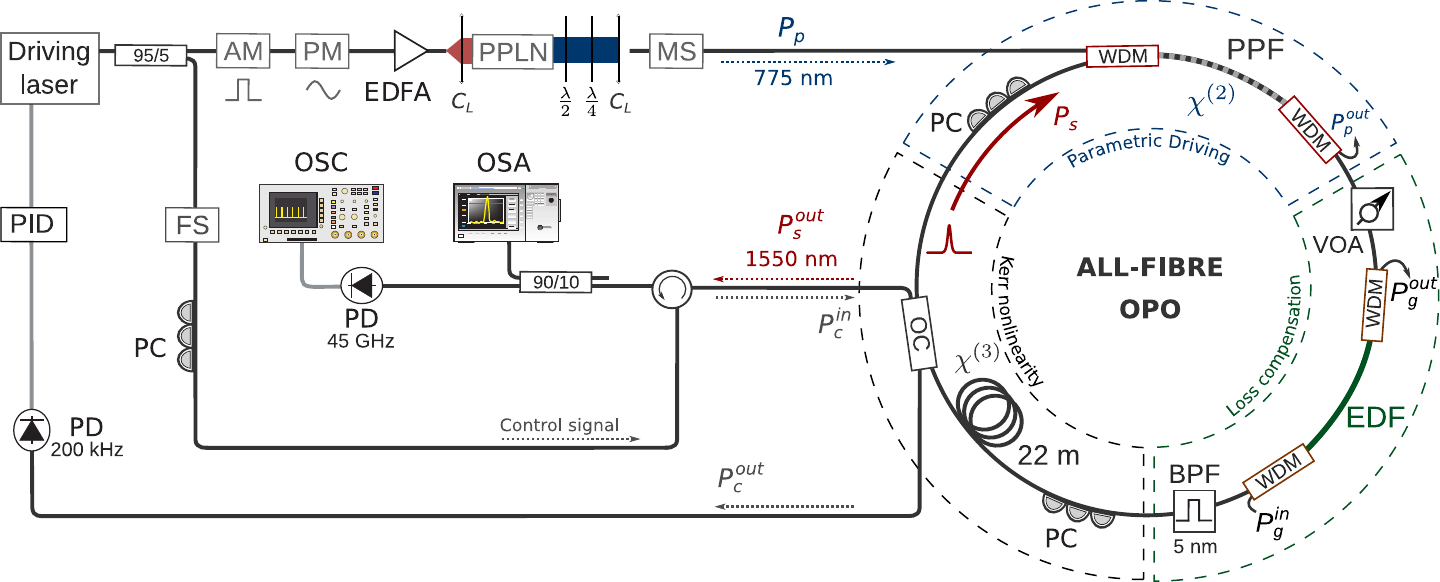}}
  \caption{\textbf{Experimental set-up.} Temporal solitons are excited in an all-fibre degenerate optical parametric oscillator (OPO). It includes a periodically poled fibre (PPF) to provide the parametric gain, a short erbium doped fibre (EDF) pumped by a 1480\,nm laser ($P_g^{in}$) for loss compensation and standard single-mode fibre. The cavity is driven by a 775\,nm pump ($P_p$), obtained by doubling the driving laser frequency in a free-space periodically poled lithium niobate (PPLN). Prior to its conversion, the driving laser is amplitude modulated (AM) and amplified through an erbium doped fibre amplifier (EDFA). A lens ($C_L$) is used to couple the light to the fibre. Two phase plates and a mode scrambler (MS) are used to limit polarization and modal losses. An output coupler (OC) is included for soliton analysis, using an optical spectrum analyzer (OSA), a fast photodiode (PD) and an oscilloscope (OSC). The cavity is actively stabilised using a proportional-integral-derivative (PID) controller and a counter-propagating beam, frequency shifted (FS) from the driving laser. PC: polarization controller;  BPF: bandpass filter; WDM:  wavelength division multiplexers; VOA: variable optical attenuator; PM: phase modulator.}
  \label{fig:setup}
\end{figure*}
\noindent We start by theoretically examining the dynamics of soliton formation in degenerate OPOs incorporating a Kerr section~\cite{longhi_ultrashort-pulse_1995}. We consider the dimensionless PNLSE~\eqref{eq:PNLSE}. The derivation of the equation and its normalization are detailed in the Supplementary Information. 
In this equation, there are only two independent parameters: the phase detuning~$\Delta$ and pump amplitude~$\mu$. They determine the 2-dimensional parameter space, plotted in Fig.~\ref{fig:Bifurcation}a, where we show the different nonlinear attractors of the system. 
The degenerate OPO threshold is located at $\mu=\sqrt{1+\Delta^2}$ and correspond to a pitchfork bifurcation (PB) of the trivial state. For negative detunings, that bifurcation is supercritical and the trivial state is modulationally unstable beyond $\mu=1$~\cite{mosca_modulation_2018}. The patterns emerging beyond this instability correspond to non degenerate oscillations, which hence has a lower threshold than degenerate emission in that region.
For positive detunings, the trivial solution is stable up to $\mu=\sqrt{1+\Delta^2}$ and the pitchfork bifurcation is subcritical. An unstable homogeneous state emerges from the trivial solution and folds at the saddle node bifurcation SN$_h$ located at $\mu=1$ (see Fig.~\ref{fig:Bifurcation}b). Beyond the fold, the upper branch is modulationally unstable, creating a region where a trivial solution and a modulated pattern coexist.
In that region ($\mu>1$), the PNLSE admits exact solitary waves of the form $u=\sqrt{2}\beta{\rm sech}(\beta t)\exp(i\phi)$ where $\cos(2\phi)=\mu^{-1}$ and $\beta^2=\Delta+\mu\sin(2\phi)$ ~\cite{miles_parametrically_1984,bondila_topography_1995,longhi_ultrashort-pulse_1995,perez-arjona_theory_2007}. There are two solitons of different amplitude and each can have one of two opposite phases. Both branches, defined as the soliton peak power, are shown in Fig.~\ref{fig:Bifurcation}b as a function of the driving power. They connect at the saddle node bifurcation SN$_s$ ($\mu=1$). The solutions corresponding to $\sin(2\phi)>0$ are always unstable.
These soliton branches are remininiscent of the ones describing CSs~\cite{scroggie_pattern_1994}.
Conversely, when plotted as a function of the detuning, see Fig.~\ref{fig:Bifurcation}c, both the homogeneous and soliton branches significantly differ from those of CSs~\cite{coen_universal_2013}. Unlike tilted resonances, the stable and saddle PCSs do not connect making the branches infinitely long.
In practice, they will be limited by higher order effects (see Supplementary Information).
Along the main soliton branch, there are a couple of Hopf bifurcations (HB).
Between these bifurcations, the PCSs are unstable and localised oscillatory behaviour as well as complex spatiotemporal dynamics can be found~\cite{bondila_topography_1995}.
In what follows, we focus on the region where stable soliton formation is predicted.\\

\noindent\textbf{Experimental setup}\\
\noindent For our experimental investigation of the PCS, we introduce an all-fibre degenerate OPO (see Fig.~\ref{fig:setup}), specifically designed so as to be governed by the PNLSE.
It is composed of three main sections made of different fibres. A 27~cm long periodically poled fibre (PPF)~\cite{lucia_thermal_2017}, a standard single mode fibre (21~m), and 52~cm of erbium doped fibre (EDF).
The first two fibres provide, separately, the quadratic and cubic nonlinearities while the EDF is used to compensate the intracavity loss~\cite{englebert_temporal_2020}.
The OPO is synchronously pumped with highly coherent 650~ps long, flat top, pulses at 775~nm. We use short pulses to keep the gain saturation low. In the region where solitons exist, our system mimicks a high finesse resonator~\cite{englebert_temporal_2020}.
The EDF is pumped with 2~W at 1480~nm. The corresponding single pass gain is 35~\%, leading to an effective finesse of 122 around 1550~nm.
The 775~nm driving signal is generated by frequency doubling a highly coherent 1550~nm laser. It is sent in the cavity through a WDM and removed after the PPF. This single pass configurations ensures that the temporal profile at the driving frequency remains nearly constant, which is crucial when aiming to observe solutions of the PNLSE (see Supplementary Information).\\

\begin{figure}
\includegraphics[scale=.67]{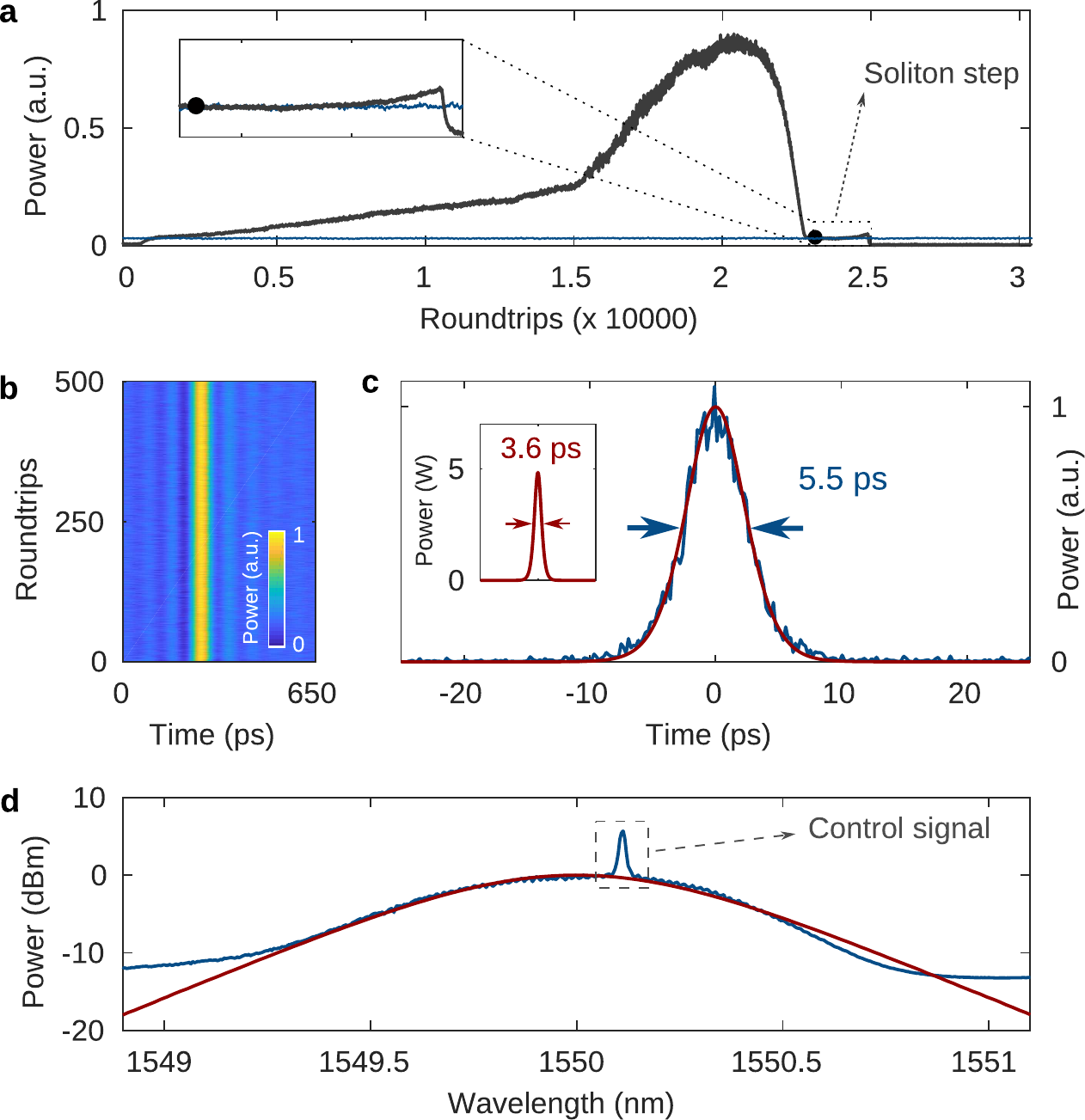}
\caption{\textbf{Characterisation of the Parametric Cavity Soliton.} {\bf a},\,Forward scan (black line) through a resonance for $P_p=10$\,W. The dot highlights the stabilisation setpoint ($\Delta=1.2$). The blue line corresponds to the output power when the cavity is actively stabilised around that level. {\bf b},\,Oscilloscope recording -- taken several seconds after the excitation process -- showing a stable, resolution limited, pulse exiting the cavity. {\bf c},\,Experimental (blue line) and theoretical (red line) autocorrelations traces. The inset shows the theoretical profile of the corresponding background-free soliton. {\bf d},\,Experimental (blue line) and theoretical (red line) spectra at the cavity output. The narrow peak corresponds to back reflections of the control signal.} 
\label{fig:soliton}
\end{figure} 

\noindent\textbf{Characterization of the PCS}\\
\noindent In a first experiment, we set the driving power to 10~W (peak), corresponding to $\mu=1.37$, and scan the laser frequency (230\,kHz/ms). Our results are shown in Fig.~\ref{fig:soliton}.
The signal resonance, measured around 1550~nm, is reminiscent of that observed in externally pumped Kerr resonators~\cite{herr_temporal_2014}. The signal average power gradually increases until it reaches the bistable region where it suddenly drops, indicating the formation of localised structures. The small plateau emerging at that point corresponds to the soliton branch shown in Fig.~\ref{fig:Bifurcation}c. In the context of externally driven Kerr resonators, it is often called the soliton step as pulses tend to merge one by one, leading to a stairs-shaped transmission curve~\cite{herr_temporal_2014}.
Additional higher resolution measurements of the nonlinear transmission of the cavity, including multi soliton steps, are shown in the Supplementary Information.
We readily note an important difference between our experimental scans and the analytical branch shown in Fig.~\ref{fig:Bifurcation}c. The soliton step in our experiments has a finite extension while the theoretical branch grows indefinitely with increasing $\Delta$. First, we stress that frequency scans are inherently dynamical such that the measured output power is not necessarily representative of steady state solutions at the corresponding
detuning. 
Second, higher order effects limit the branch in optical parametric oscillators (see Supplementary Information).
In our experiment, however, the soliton collapse is due to the 5-nm, flat top, intracavity filter we use to prevent lasing at shorter wavelengths~\cite{englebert_temporal_2020}.
As the detuning is ramped up, so is the soliton's spectral width, such that the filter eventually prevents stable soliton formation.

Next, we use a control signal to stabilise the system in the soliton region (see Methods). The average output power when the detuning is set to $\Delta=1.2$ ($\delta_0=0.03$) is shown in Fig.~\ref{fig:soliton}a. A high-resolution (80~ps) recording of the corresponding cavity output is shown in Fig.~\ref{fig:soliton}b. A resolution limited pulse can be seen exiting the cavity every rountrip time.
Further temporal (Fig.~\ref{fig:soliton}c) and spectral (Fig.~\ref{fig:soliton}d) characterisations confirm that a short (3.6~ps) pulse is circulating in the cavity. The agreement with the analytic soliton solution of the PNLSE is excellent. The experimental spectral background corresponds to the ASE emitted by the intracavity amplifier~\cite{englebert_temporal_2020}.
These measurements confirm that our novel system is governed by the PNLSE in that region and constitute, to the best of our knowledge, the first experimental observation of its well known soliton in optics.\\

\begin{figure}[H]
\includegraphics[scale=.65]{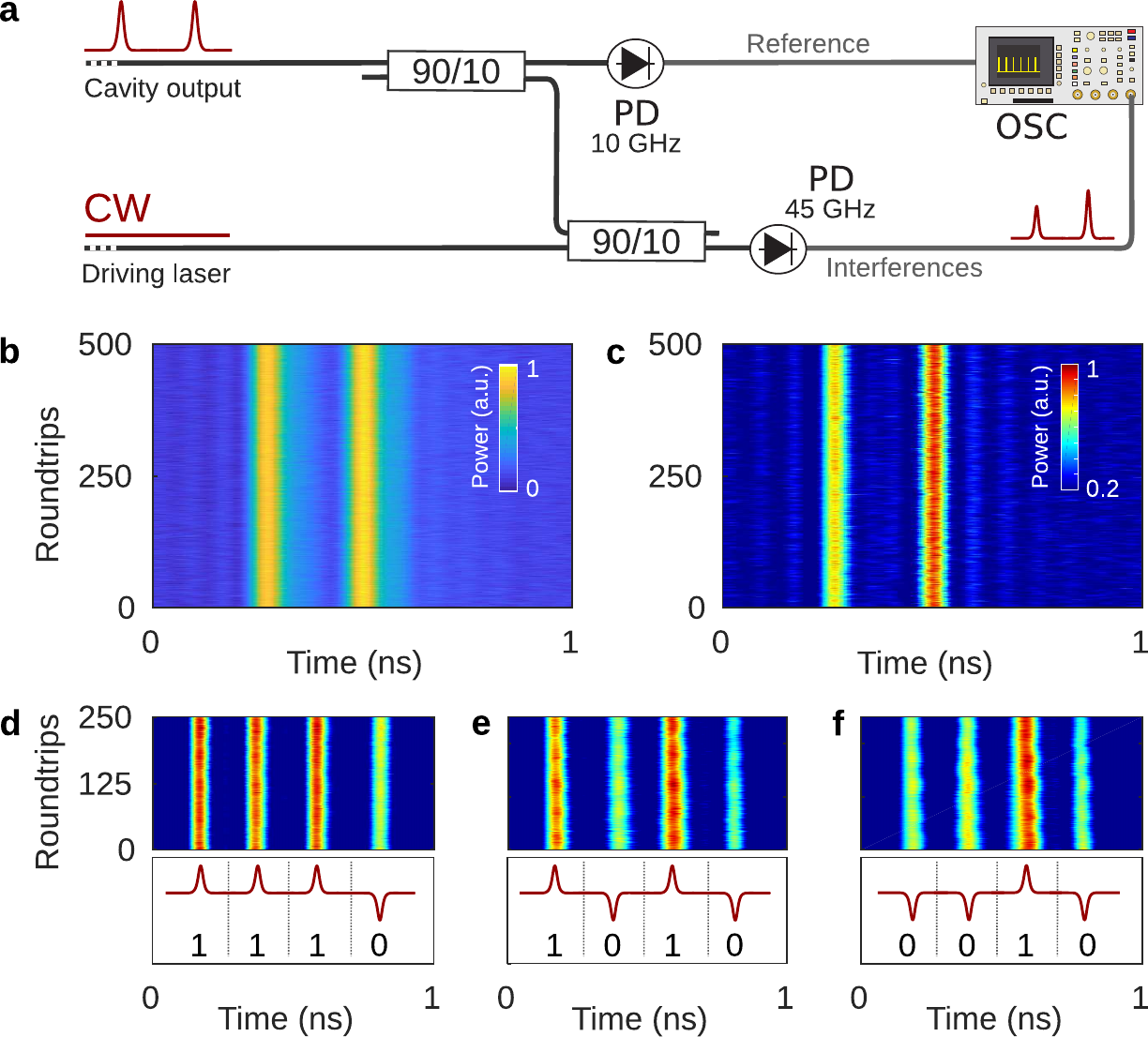}
\caption{\textbf{Random bits generation.} {\bf a},\,Experimental set-up for coherent detection. {\bf b},\,Direct detection of two PCSs. {\bf c},\,Coherent detection of two PCSs, highlighting the two different phases. {\bf d-e},\,Sequences of four random bits generated through PCS formation.} 
\label{fig:phase}
\end{figure}

\noindent\textbf{Random bits generation}\\
\noindent Parametrically driven Kerr cavity solitons are phase locked to a driving laser, as are externally driven CSs which attract a lot of attention because of their inherent stability.
The additional advantage of the PCS is its multiplicity.
Owing to the $\mathbb{Z}_2$-symmetry of the PNLSE, two attractors, which have the same amplitude but opposite phase, may coexist in the cavity, adding a degree of freedom to Kerr resonators. 
In particular, it opens the possibility to use Kerr solitons in applications, such as random bit generators~\cite{marandi_all-optical_2012} and Ising machines~\cite{inagaki_coherent_2016}, which require two different attractors.
To confirm this potential, we design a proof of principle experiment of random number generation. The concept is simple. When a soliton is spontaneously excited, it has a 50\% chance of locking to the pump with one of the two possible phase relations.
By exciting multiple solitons, and extracting the phase, we can generate a random sequence of bits. 
For this demonstration, we phase modulate the pump beam so as to excite a series of equally spaced single solitons. 
The physics behind soliton attraction to phase maxima is similar to that of CSs~\cite{jang_temporal_2015} and is detailed in the Supplementary Information. A low modulation frequency (4.6~GHz) is chosen to be able to resolve individual solitons on the oscilloscope.
We extract a portion of the 1550~nm driving laser, prior to its frequency doubling, and use it as a local oscillator for coherent detection (see Fig.~\ref{fig:phase}a). 
We excite two solitons in the cavity and send both the reference and the combined beams to a fast photodetector.
The results are shown in Fig.~\ref{fig:phase}b-c.
As expected, the reference, corresponding to the intensity, displays identical traces separated by 220~ps. After interfering with the local oscillator however, two different amplitudes are measured. These measurements confirm that solitons of different phases are excited in the cavity.
In a second series of experiments, we expand the pulse width to host four solitons and perform three distinct resonance scans. Our results are shown in Fig.~\ref{fig:phase}d-f.
By assigning a binary value to each soliton, our results correspond to a series of 4-bits random numbers, highlighting the potential of PCSs for applications. 
Moreover, our measurements confirm that the solitons are phase-locked, as only 2 distinct amplitudes are measured across the different scans.

\noindent\textbf{Discussion}\\
\noindent In summary, we investigated Kerr soliton formation in singly resonant optical parametric oscillators. We built a novel system that is well described by the seminal parametric nonlinear Schrödinger equation when driven with a frequency close to twice that of a longitudinal mode.
We theoretically showed that a couple of stable solitons exist in a broad region of experimental parameters. Our measurements confirm the existence of a backgroundless, sech-shaped and phase locked optical pulse in that region. Its temporal and spectral profiles are in excellent agreement with the soliton solution of the PNLSE. The same profile corresponds to the well known non propagating hydrodynamic soliton~\cite{miles_parametrically_1984,wu_observation_1984}.
Here, the soliton propagates along the resonator and forms an ultra-stable pulse train at the output. 
The phase-locking ensures minimal jitter and the output spectrum consists in an ultra-coherent frequency comb. Importantly, the large central peak inherent to external driving is absent.
Moreover, we showed that applications of PCSs go beyond frequency comb generation.
The two different phases can be leveraged for random number generation, as demonstrated above, or physical Ising machines. The latter has already been implemented using a synchronously pumped degenerate OPO~\cite{inagaki_coherent_2016}, but the number of individual spins is limited by the repetition rate of the pump laser. Our results show that a grid of individual spins, as dense the input phase modulation, can be generated in a long fibre cavity.
Because the number of potential connections scales as $N^2$, a 40\,GHz phase modulation would lead to a 3 orders of magnitude increase in the number of spin-spin couplings as compared to the state of the art~\cite{inagaki_coherent_2016}.

\begin{center}
    \textbf{Methods}
\end{center}
\small

\subparagraph*{\hskip-10pt Linear stability analysis}\ \\
The temporal linear stability of the steady-state solutions, shown in Fig.~\ref{fig:Bifurcation}, has been computed by solving the eigenvalue problem 
$\mathcal{L}\psi=\sigma\psi$,
obtained from the linearization of Eq.~(\ref{eq:PNLSE}) around a given steady state, where $\mathcal{L}$ the linear operator evaluated at such state,
 and $\sigma$ and $\psi$ are, respectively, the eigenvalues and eigenfunctions of $\mathcal{L}$.
This problem can be easily solved analytically for the homogeneous $u_h$ state as shown in the Supplementary Information. For the soliton state stability, we have adopted a numerical approach. We compute the eigenvalues of the Jacobian matrix obtained from $\mathcal{L}$ after spatial discretization in a $N=1024$ points grid.

\subparagraph*{\hskip-10pt Experimental set-up}\ \\
The all-fibre optical parametric oscillator (OPO) is made of a section ($L_1=27$\,cm) of periodically poled silica fibre (PPF), a section ($L_2=21$\,m) of standard telecommunication single-mode silica fibre (SMF-28) and a section ($L_3=52\,$cm) of erbium doped fibre (EDF).
The PPF has a second-order nonlinear parameter of $\kappa = 0.04\,$W$^{-1/2}$m$^{-1}$ and a phase-matching wavelength of 1548.8\,nm at room temperature. This wavelength is increased up to 1549.72\,nm to be in the tuning range of the driving laser by placing the fibre in a stabilised oven at 36\,$^\circ$C. Two wavelength division multiplexers (WDMs) are used to combine the 775\,nm pump with the intracavity signal, and to reject the remaining pump power at the fibre output.  Two different polarization controllers are used. One to align the pump polarization with the phase-matched eigenmode of the PPF and the other to align the signal polarization with one of the two eigenmodes of the cavity.
The EDF (Liekki$^{\text{®}}$ ER16-8/125) provides the optical gain. Two wavelength division multiplexers (WDMs) are inserted in the cavity to combine the 1480\,nm pump with the intracavity signal, and to reject the unabsorbed power at the amplifier output. Its length is empirically set so that the gain is slightly larger than the intrinsic cavity loss. We then use a variable optical attenuator to increase the loss and ensure the cavity is below the lasing threshold. An optical bandpass filter (5\,nm at 0.5\,dB, centred on 1550\,nm) hinders laser emission at shorter wavelengths. The cavity contains a 99/1 coupler used either to inject the control signal into the cavity or to extract part of the intracavity power.  The total intracavity loss, excluding the doped fibre, is 40\%. 
The driving continuous wave (CW) laser is a Koheras Adjustik$^{\text{TM}}$ E15 with a sub-100~Hz linewidth. Its wavelength is set to 1549.72\,nm, on the edge of the tuning range (1\,nm) to coincide with the PPF phase-matching wavelength. The laser output is first modulated with a Mach-Zehnder amplitude modulator (bandwidth: 12 GHz, extinction ratio: 30\,dB), driven by a pattern generator connected to an RF clock. The pulsed beam is then amplified with an erbium doped fibre amplifier (EDFA) and converted to its second harmonic through a 4\,cm long periodically poled lithium niobate (Covesion$^{\text{®}}$ MSHG1550-0.5-40, $\kappa = 2.5\,$W$^{-1/2}$m$^{-1}$) in a free-space section. Using dichroic mirrors, the unconverted field is attenuated by 125\,dB such that only the pump is injected into the fibre. To minimize both polarization and modal losses at the first WDM, a half- and quarter-wave plate (free-space) and a mode scrambler are used. 
The cavity resonances are measured by scanning the frequency of the driving laser and recording the average power at the output coupler, $P_s^{out}$, with a 200\,kHz photodiode (see e.g. \autoref{fig:soliton}a). To stabilise the cavity, a control signal is generated by extracting a portion of the driving laser power through a 95/5 coupler and shifting its frequency with a tunable frequency-shifter ($ 110\pm5\,$MHz). 
Using a circulator and a polarization controller, the counter-propagative control signal, with a power of $P_c^{in}$, is sent to the cavity on the orthogonal polarization eigenmode to avoid seeding the OPO. The cavity detuning is stabilised by slightly changing the driving laser wavelength to maintain a constant control signal output power $P_c^{out}$. The feedback signal is generated by a proportional-integral-derivative (PID) controller (Toptica DigiLock 110), driven by a photodiode. The system is stabilised on the slope of the linear resonance of the control signal. Knowing the cavity birefringence, the detuning of the signal can be extracted. It can then be modified by changing the control signal frequency~\cite{li_experimental_2020}. 
Part of the intracavity power is extracted at the output coupler to characterize the solitons. The spectrum of the parametric cavity soliton (PCS) is recorded on an optical spectrum analyser
(0.1\,nm resolution bandwidth). Time measurements are carried out with a fast photodiode (45\,GHz bandwidth) and an oscilloscope (10\,GHz bandwidth, 10 Gsample.s$^{-1}$). The intensity autocorrelation trace is directly acquired at the cavity output. For this measurement, a commercial EDFA is used to increase the average output power to $\sim$70\,mW.

\subparagraph*{\hskip-10pt PPF fabrication}\ \\
  The PPF is a 125\,$\mu$m outside diameter cladding fibre with a Germania-doped glass core of 4\,$\mu$m diameter and a numerical aperture NA\,$=0.17$. Two 27\,$\mu$m diameter channels run adjacent to the fibre core at a distance of respectively 13.6\,$\mu$m and 7.2\,$\mu$m from the core’s edges. The fibre is first thermally poled in single anode configuration at 265$^\circ$C with an electric potential of $+8\,$kV applied to the embedded electrode, for 2 hours~\cite{lucia_single_2019}. The second order nonlinearity created via thermal poling is then erased periodically by means of a CW argon ion laser frequency doubled to 244\,nm, equipped with an acousto-optic modulator (AOM) used to modulate the laser output. The laser is focused to a circular spot, 20\,$\mu$m in diameter, while the poled fibre is clamped onto a linear stage by two fibre rotator clamps. The laser is modulated using the AOM while translating the fibre core through the spot to achieve a grating of the desired duty cycle and period. For the grating a fluence of 14\,J/cm$^2$ and a duty cycle of 5\% was used to periodically erase the nonlinearity. The period of the grating was chosen to be 55\,$\mu$m in order to have quasi-phase matching at a wavelength around 1550 nm.\\
  
\subparagraph*{\hskip-10pt PCS excitation and stabilisation}\ \\
For all measurements depicted in Fig\,4, the cavity is synchronously pumped with 650\,ps flat-top pulses whose repetition frequency matches the cavity free-spectral range (FSR). The cavity detuning is stabilised by locking the control signal through-port transmission at 90\% (i.e. $P_c^{out}/P_c^{in}=0.9$). Once stabilised, the control signal frequency is increased with the frequency-shifter (FS) until the signal output power drops to the soliton-step power (Fig.\,4a, red). This coincides with the emergence of a background-free sech squared-like spectrum, corresponding to the generation of a single PCS. Temporal and spectral measurements are then carried out. 

\subparagraph*{\hskip-10pt Coherent detection measurement}\ \\
To demonstrate the existence of PCS with opposite phases, the cavity is synchronously pumped with 1\,ns or 1.9\,ns flat-top pulses. On these pump pulses, we also imprint a 4.6\,GHz phase modulation (PM) using a phase-modulator. As for CSs~\cite{jang_temporal_2015,jang_controlled_2016}, PCS are attracted by PM maxima (see Supplementary Information). When scanning the resonance, we generate up to four PCSs, separated by 220\,ps. Using a 90/10 coupler, most of the cavity output power $P_s^{out}$ is sent to a 10\,GHz photodiode [i.e. reference beam on Fig.\,5a]. The remaining power is combined with part of the driving laser power, obtained by bypassing the frequency-shifter, through another 90/10 coupler. The result of the interference is sent to a 45\,GHz photodiode for coherent detection. \\

\begin{center}
    \textbf{Acknowledgements}
\end{center}
\noindent We are grateful to Michaël Fita Codina for the manufacturing of experimental components and
to Pascal Kockaert and Costantino Corbari for fruitfull discussions. This work was supported by funding from the European Research Council (ERC) under the European Union’s Horizon 2020 research and innovation programme (grant agreement No 757800). 
N.E. acknowledges the support of the Fonds pour la formation à la Recherche dans
l’Industrie et dans l’Agriculture (FRIA, Belgium). P.P.R. acknowledges the support of the "Fonds de la Recherche Scientifique" (FNRS, Belgium).\\

\begin{center}
    \textbf{Author Contributions}
\end{center}
\noindent N.E. designed and performed the experiments, supervised by S-P.G. F.D.L. and P.J.S. manufactured the periodically poled fibre. NE derived and simulated the mean-field model. P.P.R. and C.M.A. performed the bifurcation and linear stability analysis of the mean-field model. F.L. supervised the overall project and wrote the manuscript. All authors discussed the results and contributed to the final manuscript.\\

    \bibliography{Article}
	\bibliographystyle{naturemag}

\newpage\ 
\newpage
\includepdf[pages=1]{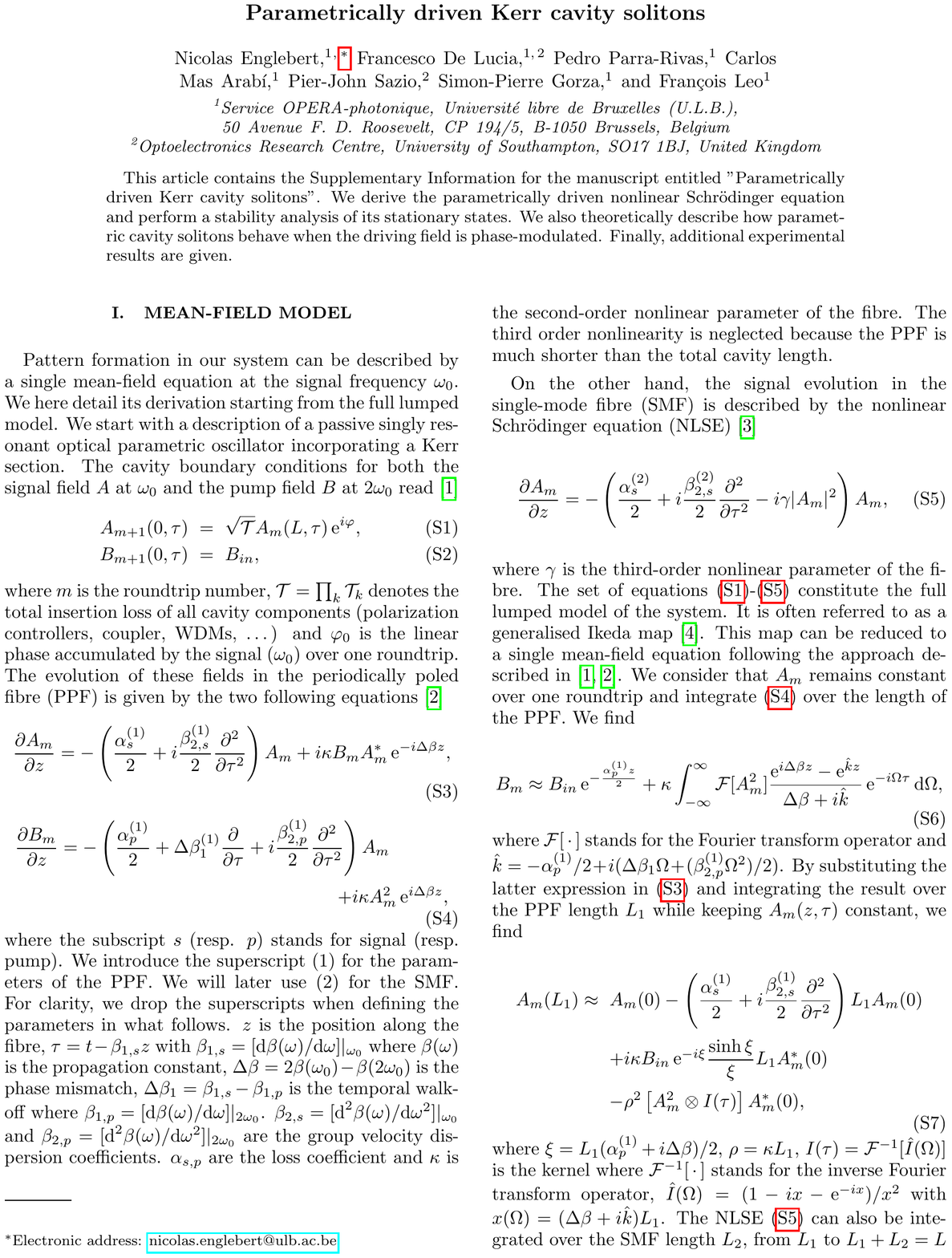}\newpage\ \newpage 
\includepdf[pages=2]{SI.pdf}\newpage\ \newpage
\includepdf[pages=3]{SI.pdf}\newpage\ \newpage
\includepdf[pages=4]{SI.pdf}\newpage\ \newpage
\includepdf[pages=5]{SI.pdf} 
	
\end{document}